\documentclass{Interspeech2024}
\usepackage{multirow}
\usepackage{multicol}
\usepackage{amsmath}
\usepackage{amssymb}

\interspeechcameraready

\title{SCDNet: Self-supervised Learning Feature based Speaker Change Detection }

\name[affiliation={}]{Yue}{Li}
\name[affiliation={}]{Xinsheng}{Wang}
\name[affiliation={}]{Li}{Zhang}
\name[affiliation={*}]{Lei}{Xie}

\address{
  Audio, Speech and Language Processing Group (ASLP@NPU), School of Software, \\ 
  Northwestern Polytechnical University, Xi'an, China
}
\email{yueli77@mail.nwpu.edu.cn, lxie@nwpu.edu.cn \thanks{* Corresponding author.}}


\begin{document}

\maketitle

\begin{abstract}
    Speaker Change Detection (SCD) is to identify boundaries among speakers in a conversation. Motivated by the success of fine-tuning wav2vec 2.0 models for the SCD task, a further investigation of self-supervised learning (SSL) features for SCD is conducted in this work. Specifically, an SCD model, named SCDNet, is proposed. With this model, various state-of-the-art SSL models, including Hubert, wav2vec 2.0, and WavLm are investigated. To discern the most potent layer of SSL models for SCD, a learnable weighting method is employed to analyze the effectiveness of intermediate representations. Additionally, a fine-tuning-based approach is also implemented to further compare the characteristics of SSL models in the SCD task. Furthermore, a contrastive learning method is proposed to mitigate the overfitting tendencies in the training of both the fine-tuning-based method and SCDNet. Experiments showcase the superiority of WavLm in the SCD task and also demonstrate the good design of SCDNet.

\end{abstract}

\vspace{-0.05cm}
\section{Introduction}

Speaker Diarization (SD), a pivotal method in speech processing, aims to answer the question of `who speaks when' in scenarios involving multiple speakers~\cite{hruz2017convolutional}. In contrast, Speaker Change Detection (SCD) is to find the speaker turn points in the conversation~\cite{kunevsova2023multitask}, and thus it can be regarded as a subtask of SD~\cite{yin2018neural}, and also with broad applications, e.g., enhancing Automatic Speech Recognition (ASR) accuracy~\cite{sari2020auxiliary} and syncopate captioning~\cite{donabauer2021making}. 

The metric-based approach is a common early method for the SCD task, wherein speaker change points are identified through the comparison of distributions between two consecutive speech windows~\cite{kemp2000strategies}. Following the emergence of i-vector~\cite{dehak2010front} and DNN-based embeddings~\cite{snyder2018x}, uniform segmentation schemes have gained popularity as effective methods~\cite{sell2018diarization}. In this approach, the target audio undergoes segmentation into a series of segments with a constant window length and overlap length. Subsequently, speech embeddings from various segments are compared to determine if the speaker has changed. However, due to the fixed window length, a trade-off is inevitable between the efficacy of speech embedding and the accuracy of boundary detection.

To overcome the limitations of segment-based methods, various works have endeavored to predict speaker change points at the frame level through neural networks~\cite{hruz2018lstm, hruz2017convolutional, fan2022sequence}. In these approaches,  the model is generally trained with ground-truth SCD labels to perform a binary classification task. To be specific, in~\cite{hruz2018lstm}, using LSTM as the backbone, the optimizing target is to minimize the distance between the predicted probability signal and linear fuzzy labeling signal.  


In addition to label-based methods for frame-level SCD, several works have explored leveraging text transcription for word-level speaker change detection through ASR techniques~\cite{xia2022turn,zhao2023augmenting}. For example, in~\cite{xia2022turn}, the transcription used to train an ASR model is enhanced by incorporating a distinct token designed to denote speaker turns. Then the augmented transcription is used to train an ASR model that predicts not only regular text tokens but also special speaker turn tokens. While this approach alleviates the necessity for boundary annotations, using the textual transcription can be more intricate, especially in a dialogue scenario characterized by frequent interruptions and insertions, and the prevalence of intonation markers. 
Additionally, because the predicted boundaries in this method operate at the word level, the precision of boundary predictions may not be as high as those based on frame-level predictions.

Most recently, Kune{\v{s}}ov{\'a} and Zaj{\'\i}~\cite{kunevsova2023multitask} explored the effectiveness of one of the most popular SSL models, wav2vec 2.0~\cite{baevski2020wav2vec}, on the SCD task. In their research, the pre-trained wav2vec 2.0 is fine-tuned in an end-to-end way involving multi-tasks, i.e., SCD, Overlapping Speech Detection (OSD), and Voice Activity Detection (VAD). This wav2vec 2.0 and multitask-based method showcases the remarkable performance, achieving a state-of-the-art (SOTA) level in the SCD task. Inspired by this research, we are conducting a further investigation into SSL-based end-to-end training methods for SCD.

On the one hand, due to the typically large number of parameters in SSL models, directly fine-tuning them requires a certain threshold of data and computational resources. On the other hand, despite efforts by Kune{\v{s}}ov{\'a} and Zaj{\'\i}~\cite{kunevsova2023multitask} to enhance SCD performance through multitasking, e.g., OSD and VAD, all these tasks are frame-level binary classification tasks, which pose a risk of overfitting when training complex models due to the simplistic learning paradigm. Additionally, besides wav2vec 2.0, other SSL models such as Hubert~\cite{hsu2021hubert} and WavLm~\cite{chen2022wavlm} have also gained significant attention in various downstream tasks, such as Hubert-based speech recognition~\cite{nasersharif2023speech} and WavLm-based speech synthesis~\cite{lajszczak2024base}. However, the performance of these models in SCD has not been explored.  

To tackle those issues, we propose an innovative end-to-end SCD model, referred to as SCDNet, based on the Conformer architecture~\cite{gulati2020conformer}. SCDNet leverages off-the-shelf features as inputs and undergoes end-to-end training to accomplish the SCD task. Additionally, we propose a contrastive learning method for training SCD-oriented models to address the overfitting tendency associated with the frame-level binary classification task. Furthermore, we explore the performance of various SSL features via both SCDNet and fine-tuning-based methods.

\section{Approach}
SCDNet is a Conformer-based model to achieve the frame-level binary classification with speech representation as input. In addition to the classification loss, a contrastive loss is proposed to alleviate the overfitting tendency caused by the simplistic binary learning way. 
This contrastive loss is also used to fine-tune the pre-trained SSL models for the SCD task. 

\subsection{Problem Formulation}

A speaker change point is defined as the point indicating the initiation or conclusion of an individual's speech, regardless of the presence or absence of other speakers. Therefore, the consideration extends beyond transitions between two speakers to encompass voice activity boundaries. Following~\cite{kunevsova2023multitask}, SCD is treated here as a frame-level classification task. Given a speech feature sequence $X=\{x_0,x_1,...,x_T\}$ and the corresponding label sequence $Y=\{y_0,y_1,...,y_T\}$ where $T$ denotes the total number of frames, and $y_i\subseteq \{0,1\}$. For a model $f$ with learnable parameters $\theta$, the training target of SCD is formulated as:
\begin{equation}
    f_\theta = \arg\max_\theta P(Y; X, \theta)
\end{equation}

\subsection{SCDNet}

\begin{figure}[t]
  \centering
  \includegraphics[width=\linewidth]{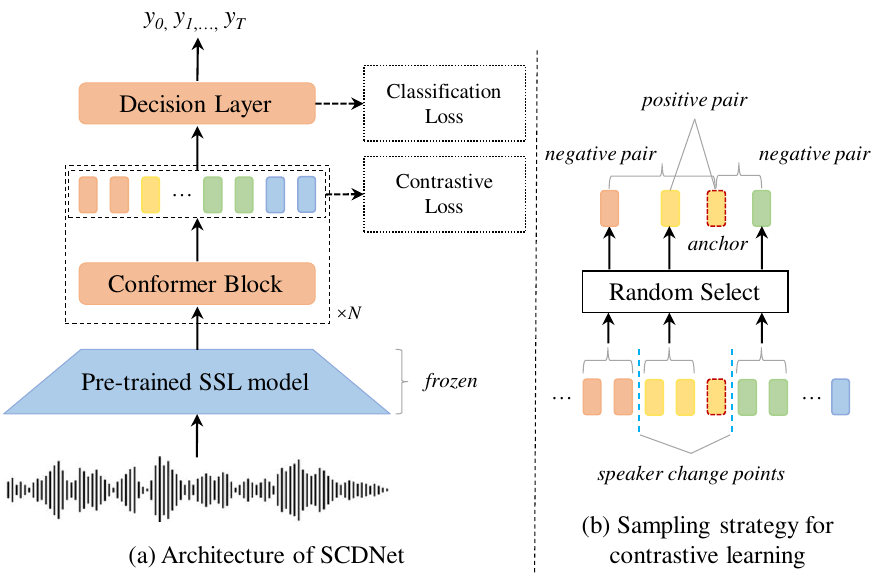}
  \caption{The architecture of the SCDNet (left) and sampling strategy for contrastive learning (right).}
  \label{fig:system_overview}
  \vspace{-0.4cm}
\end{figure}

As depicted in Figure~\ref{fig:system_overview}, the proposed SCDNet primarily comprises three components: the pre-trained SSL model, the Conformer Blocks, and the Decision Layer. During the inference process, the input audio is represented by features extracted from the pre-trained SSL model. Subsequently, these features pass through $N$-layer Conformer Blocks before producing the final boundary labels through the Decision Layer.

As a frame-level binary classification task, the classification loss, e.g., cross-entropy loss or distance-based loss, is the typical loss function for the training of SCD-related models. However, relying solely on classification loss for training can be challenging due to the limited information provided by binary labels, making it susceptible to overfitting. To address this challenge, a contrastive learning method is proposed for training SCDNet associated with classification loss.

\textbf{\textit{The classification loss}} is the basic loss function for the binary classification task. 
Considering the potential errors introduced by manual labeling, the boundaries annotated by humans may exhibit a shift from the actual boundaries. Therefore, following~\cite{kunevsova2023multitask}, instead of using the original hard label, i.e., 0 or 1, a fuzzy labeling strategy is employed. Specifically, in the original label sequence $Y=\{y_0,y_1,...,y_T\}$, $y_i=1$ means the speaker change point, and the points between two change points are all zeros. Here, with the fuzzing strategy, the label value decreases to zero from the change point linearly within 0.2s. Labels that are more than 0.2s away from the nearest change point are set to zero. With updated label $y_i$, the loss function for the classification is given by:
\begin{equation}
    \mathcal{L}_p = \frac{1}{T} \sum_{i=1}^{T} || \hat{y}_i - y_i ||,
\end{equation}
where $\hat{y}_i$ is predicted value.

\textbf{\textit{Contrastive learning}}, which yields the contrastive loss, aims to ensure the distinctiveness of representations generated by each Conformer block layer. This serves to mitigate the risk of overfitting during the training of the SCD model. The fundamental concept is to make representations between two change points distinctive from those of adjacent regions. Hence, contrastive learning for SCD aims to enhance the similarity of representations within the same segment while diminishing the similarity with representations in adjacent segments. Here, a segment refers to the region between two speaker change boundaries. 

As illustrated on the right side of Figure~\ref{fig:system_overview}, given a frame-level representation $ h_{i}^{j}$ as the anchor, where $i$ means the position index of the representation sequence and $j$ means the layer index from $N$ Conformer block layers, the positive sample $h_{p}^{j}$, is randomly chosen from the same segment. Simultaneously, the negative sample $h_{n}^{j}$ is randomly selected from one adjacent segment, either on the right one or on the left one, or a randomly sampled vector if no adjacent segment exists.

Based on the anchor $ h_{i}^{j}$, the positive sample $h_{p}^{j}$, and the negative sample $h_{n}^{j}$, the contrastive loss is defined as:

\begin{equation}
\begin{aligned}
    \mathcal{L}_c &= - \frac{1}{T \cdot N} \sum_{i=1}^{T}\sum_{j=1}^{N}(\log S( h_{i}^{j}, h_{p}^{j}) \\
    &\quad \quad \quad \quad \quad + \log [1 - S( h_{i}^{j}, h_{n}^{j})]),
\end{aligned}
\end{equation}
where $S$ is to calculate the cosine similarity between two frame-level features and is given by 
\begin{equation}
    S(h_i, h_p) = \frac{h_i \cdot h_p} {||h_i|| \cdot ||h_p||}
\end{equation}

\textbf{\textit{The total loss }} is calculated by:
\begin{equation}
    \mathcal{L} = \mathcal{L}_p + \alpha \mathcal{L}_c
\label{eq:loss}
\end{equation}
where $\alpha$ is a hyper-parameter to balance the weight between $\mathcal{L}_p $ and $\mathcal{L}_c$.

\subsection{SSL Features for SCDNet}
The intermediate representations from different layers of the same pre-trained SSL model typically exhibit distinct properties~\cite{pasad2023comparative}. Hence, directly utilizing features from the last layer may not be optimal. To effectively identify the most influential layer for the SCD task, a weighting fusion strategy is employed to assess the contribution of each layer's representation. To be specific, for an SSL model with $L$ layers, the representation from layer $l$ is denoted as $X_l$, and the fused representation is obtained as follows:

\begin{equation}
    X = \sum_{l=1}^{L} \sigma_l X_l
\label{eq:weight}
\end{equation}
where $\sigma_l$ is a learnable parameter that weights the representation from layer $l$. Following the completion of training, a larger $\sigma_l$ suggests a greater contribution from the corresponding layer. This information can be utilized to identify the most influential layer for extracting representations in the SCD task.

\subsection{Fine-tuning SSL Models for SCD}

In addition to the off-the-shelf representation-based SCDNet, we also assess the performance of various SSL models in the SCD task through fine-tuning. Following the methodology outlined in~\cite{kunevsova2023multitask}, only the parameters from the transformer layers and the decision layer are updated during the fine-tuning process. However, unlike~\cite{kunevsova2023multitask}, where the fine-tuning employs a multi-task loss function, in this study, the loss function is based on Eq.~\ref{eq:loss}.

This fine-tuning approach serves a dual purpose: it compares the performance of fine-tuning different SSL models in the SCD task and enables a direct comparison between the multi-task-based loss in~\cite{kunevsova2023multitask} and the proposed loss function.

\section{Experimental settings}

\subsection{Dataset and Evaluation}
Four real datasets, including AMI~\cite{kraaij2005ami, carletta2005ami}, AliMeeting~\cite{yu2022m2met}, AISHELL-4~\cite{fu2021aishell}, and 
DIHARD3~\cite{ryant2020third}, are used to evaluate the proposed method. For the AMI dataset, the ``headset mix" recordings are utilized. The far channel 0 and channel 0 of AliMeeting and AISHELL-4 are adopted, respectively. 
In addition to the above real datasets, an artificial dataset is created from the ``train-other-500" subset of LibriSpeech~\cite{panayotov2015librispeech} based on the simulation procedure described in \cite{fujita2019end}. 

Considering the widespread use of AMI in the SCD task, the comparison with other methods is performed on the AMI dataset, while other datasets are used to further validate the robustness of SCDNet and demonstrate the effectiveness of the contrastive learning method.

Following~\cite{kunevsova2023multitask}, purity~(Pur) and coverage~(Cov) scores~\cite{yin2017speaker} are adopted as the evaluation metric for the SCD task, and F1 presents the harmonic mean of these two. The Python library \textit{pyannote.metrics}\footnote{Downloaded from \href{https://pyannote.github.io/}{https://pyannote.github.io/}} \cite{bredin2017pyannote} is used to compute the corresponding metric.

\subsection{Implementation details}

The SCDNet comprises a 3-layer Conformer block ($N=3$) with a hidden dimension of 384. The parameter $\alpha$ in Eq.~\ref{eq:loss} is set as 0.05. During inference, a threshold of 0.35 is employed to binarize the predicted probabilities of speaker change points generated by the model.

\section{Experimental Results}

\subsection{SSL Representation Comparison}
Various recently popular SSL models, including wav2vec 2.0, Hubert, WavLm, and their different scales are taken into consideration, which can be found in Table~\ref{tb:ssl_detail}. Both SCDNet-based and fine-tuning-based methods are employed to explore the effectiveness of these models in the SCD task.

\begin{table}[]
\caption{The details of SSL models' parameters and pre-training data.}
\vspace{-0.2cm}
\setlength{\tabcolsep}{1.95mm}
\begin{tabular}{lcc}
\toprule
Model                                                 &  Parameters(M)   & Data     \\ 
\midrule
hubert-base-ls960 \cite{hsu2021hubert}                &  95          & LS-960   \\ 
wav2vec2-base \cite{baevski2020wav2vec}               &  95          & LS-960   \\
wavlm-base \cite{chen2022wavlm}                       &  95          & LS-960   \\
hubert-large-ll60k \cite{hsu2021hubert}               &  317         & LL-60k   \\ 
wav2vec2-large-xlsr-53 \cite{baevski2020wav2vec}      &  317         & LL-60k   \\
wavlm-large \cite{chen2022wavlm}                      &  317         & MIX-94k  \\
\midrule
SCDNet                    &  33             & - \\ \bottomrule
\end{tabular}
\label{tb:ssl_detail}
\vspace{-0.2cm}
\end{table}

\begin{figure}[t]
  \centering
  \includegraphics[width=\linewidth]{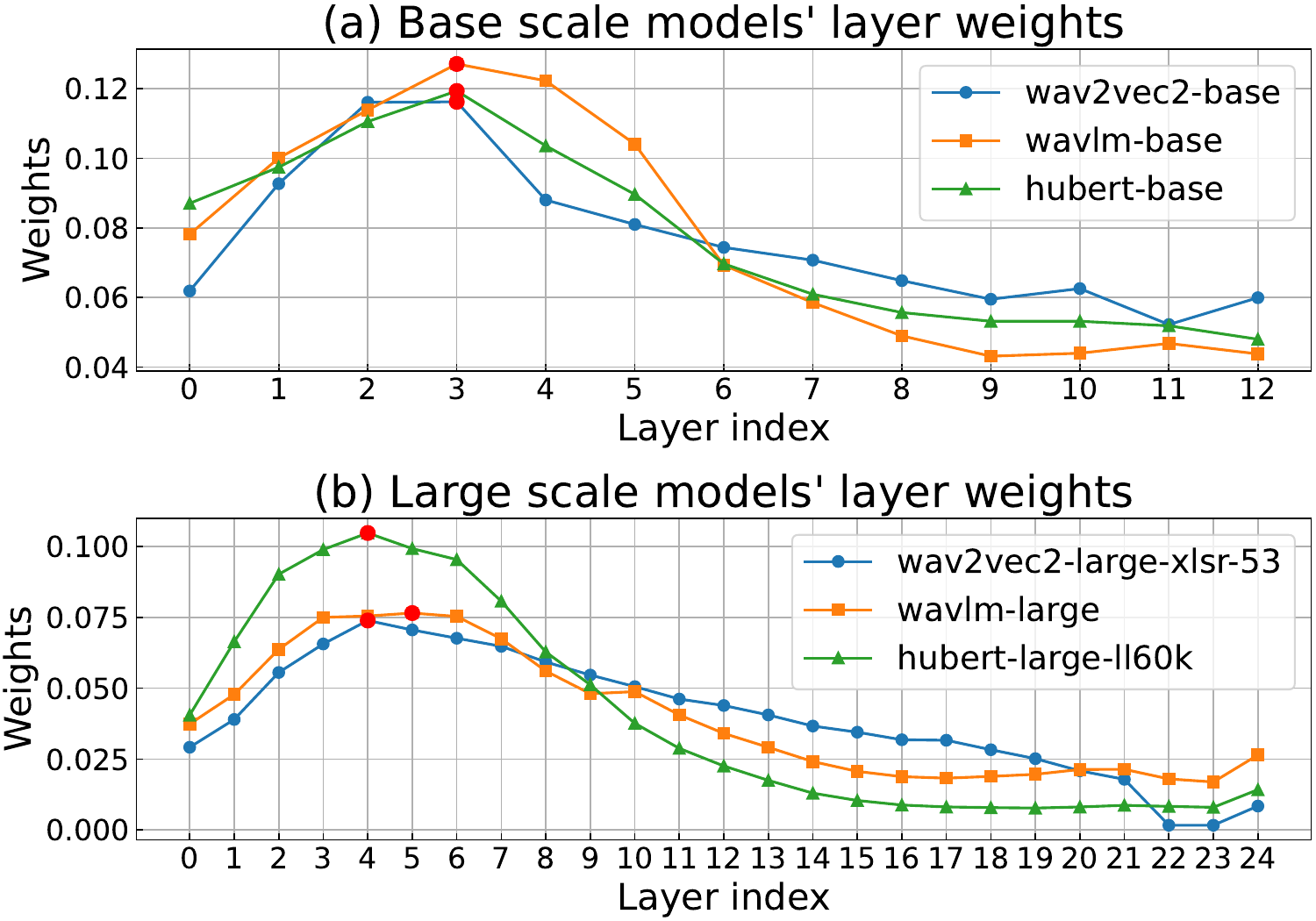}
  \setlength{\abovecaptionskip}{-0.2cm}
  \caption{Weighting values of different layers in the weighted representation fusion method. Experiments are conducted using the AMI dataset.}
  \label{fig:layer_weight}
  \vspace{-0.5cm}
\end{figure}

\textbf{\textit{The SCDNet-based SSL exploration}} is initiated with the weighting fusion strategy to examine which layer's representation from a given SSL model is most influential in the SCD task. Figure~\ref{fig:layer_weight} illustrates the learnable weighting values ($\sigma_l$ in Eq.~\ref{eq:weight}) corresponding to different transformer layers ($l$) of an SSL model. A higher value for a layer indicates that the representation from this particular layer contributes more significantly to the final representation, in the context of the SCD task.

As depicted in Figure~\ref{fig:layer_weight}, the weighting values from different models, irrespective of whether they are base or large models, exhibit a similar trend. Specifically, these values increase from the initial layer to a certain layer and then gradually decrease. This trend aligns with the observation in~\cite{pasad2023comparative}, which suggests that the initial layers contain more acoustic information, while the deeper layers contain more semantic information. In the SCD task, both acoustic features and semantic information are valuable. The intermediate layers, striking a balance between acoustic and semantic information, demonstrate more significant contributions than the representations from the two ends.

\begin{table}[]
\caption{SCDNet performance based on various SSL features on AMI dataset.}
\setlength{\tabcolsep}{1.9mm}
\begin{tabular}{llcccc}
\toprule
Model                        & Scale                  & Layer & Cov(\%) & Pur(\%) & F1(\%)                        \\ \midrule
\multirow{4}{*}{Hubert}      & \multirow{2}{*}{base}  & 3     & 94.46 & 91.62 & 93.01                     \\
                             &                        & 12    & 91.97 & 90.73 & 91.35                     \\ \cline{2-6} 
                             & \multirow{2}{*}{large} & 4     & 94.28 & 91.68 & \rule{0pt}{2.5ex}92.96                     \\
                             &                        & 24    & 96.71 & 85.84 & 90.95                     \\ \hline
\multirow{4}{*}{wav2vec 2.0} & \multirow{2}{*}{base}  & 3     & 92.96 & 92.13 & \rule{0pt}{2.5ex}92.55                     \\
                             &                        & 12    & 92.16 & 91.44 & 91.80                     \\ \cline{2-6} 
                             & \multirow{2}{*}{large} & 4     & 93.69 & 92.86 & \rule{0pt}{2.5ex}\textbf{93.27}                     \\
                             &                        & 24    & 94.79 & 67.18 & 78.63                     \\ \hline
\multirow{4}{*}{WavLm}       & \multirow{2}{*}{base}  & 3     & 93.72 & 92.35 & \rule{0pt}{2.5ex}\textbf{93.03}                     \\
                             &                        & 12    & 91.72 & 90.37 & 91.04                     \\ \cline{2-6} 
                             & \multirow{2}{*}{large} & 5     & 94.58 & 91.50 & \rule{0pt}{2.5ex}93.01                     \\
                             &                        & 24    & 94.22 & 91.62 & 92.91 \\ \bottomrule
\end{tabular}
\label{tab:ssl_scdnet}
\end{table}

The performance of representations from the layer with the highest weighting value and the last layer is summarized in Table~\ref{tab:ssl_scdnet}. It is evident that, for each model, the intermediate representation with the highest weighting value outperforms that achieved by the last layer. This underscores the effectiveness of the weighting fusion method in identifying the influential layer, as opposed to directly utilizing the last layer. Comparing all the results, although the representation from layer 3 of Wavlm-base is inferior to the best value achieved by layer 4 of wav2vec 2.0-large, its smaller model scale and less obvious performance disadvantage make it more suitable for SCDNet. 

\textbf{\textit{The fine-tuning-based SSL comparison}} for the SCD is presented in Table~\ref{tb:ft}. As can be seen, WavLm-based methods both with large scale and base scale achieve the best performance compared with other SSL models with a similar scale, indicating that WavLm is particularly well-suited for the SCD task.

\begin{table}[]
\caption{SCD performance by fine-tuning various SSL models on AMI dataset.}
\setlength{\tabcolsep}{2.87mm}
\begin{tabular}{llccc}
\toprule
Model       & Scale                  & Cov(\%) & Pur(\%) & F1(\%) \\ \midrule
Hubert      & \multirow{3}{*}{base}  & 92.82 & 93.00 & 92.91   \\
wav2vec 2.0 &                        & 92.19 & 93.48 & 92.83  \\
WavLm       &                        & 93.43 & 93.60 & \textbf{93.51}   \\ \midrule
Hubert      & \multirow{3}{*}{large} & 93.17 & 93.20 & 93.18   \\
wav2vec 2.0 &                        & 91.63 & 93.34 & 92.47   \\
WavLm       &                        & 94.11 & 94.63 & \textbf{94.37}  \\ \bottomrule
\end{tabular}
\label{tb:ft}
\vspace{-0.3cm}
\end{table}

\subsection{Comparison with SOTA methods}

The comparison of the proposed SCDNet with previous methods is presented in Table~\ref{tab:compare_with_others}. In this table, SCDNet refers to the proposed model with the representation from layer 3 of WavLm-base as input, and it will be the default setting hereafter unless specifically mentioned otherwise. It is evident that SCDNet achieves the best performance, with a relative gain of 2.5\% in terms of F1 compared to the previous SOTA performance achieved by \cite{kunevsova2023multitask}. This result highlights the effectiveness of the design of SCDNet. 

It is noteworthy that the previous SOTA performance in~\cite{kunevsova2023multitask} is based on fine-tuning the wav2vec2-base model, the same model as presented in Table~\ref{tb:ft}. However, our results achieved by fine-tuning the wav2vec-base with the proposed contrastive learning method are notably superior to those in \cite{kunevsova2023multitask}. This superiority underscores the effectiveness of the proposed contrastive learning approach. Further evidence of this superiority will be explored in the following ablation study.

\begin{table}[th]
  \caption{Comparison of the proposed scheme with previously reported results for the SCD task on AMI dataset.}
  \setlength{\tabcolsep}{3.3mm}
  \label{tab:compare_with_others}
  \centering
  \begin{tabular}{lccc}
    \toprule
    Method & Cov(\%) & Pur(\%) & F1(\%)  \\
    \midrule
   Kune{\v{s}}ov{\'a} \textit{et al}. \cite{kunevsova2023multitask}     & 91.68 & 89.91 & 90.79        \\
    Su \textit{et al}. \cite{su2022multitask}                        & 91.75 & 85.68 & 88.61        \\
    Fan \textit{et al}. \cite{fan2022sequence}                       & 89.81 & 83.92 & 86.76        \\
    pyannote \cite{bredin2020pyannote}                            & 84.20 & 90.40 & 87.19        \\
    \midrule
    SCDNet               & 93.72 & 92.35 & \textbf{93.03}      \\
    \bottomrule
  \end{tabular}
\end{table}

\subsection{Ablation Study}

To demonstrate the effectiveness of the proposed contrastive learning method, ablation experiments were conducted on more datasets, and the corresponding results are presented in Table~\ref{tab:ablation}. All F1 values achieved by the model trained without contrastive learning are lower than those with contrastive learning on the same database. Specifically, a relatively 3.4\% higher value is observed when contrastive learning is adopted compared to that without contrastive learning on the AI-SHELL-4 database. These results collectively demonstrate the efficacy of the proposed contrastive learning in enhancing the performance of SCDNet.

\begin{table}[]
\caption{ Results of SCDNet w/wo contrastive learning (CL).}
\vspace{-0.2cm}
\label{tab:ablation}
\setlength{\tabcolsep}{3.1mm}
\begin{tabular}{lcccc}
\toprule
Dataset                     & CL   & \multicolumn{1}{c}{Cov(\%)} & \multicolumn{1}{c}{Pur(\%)} & \multicolumn{1}{c}{F1(\%)} \\ \midrule
\multirow{2}{*}{AMI}        & $\checkmark$    & 93.72 & 92.35 & \textbf{93.03}       \\
                            & $\times$  & 89.25 & 94.03 & 91.57       \\ \hline
\multirow{2}{*}{AliMeeting} & $\checkmark$    & 93.57 & 86.61 & \rule{0pt}{2.4ex}\textbf{89.95}       \\
                            & $\times$  & 94.52 & 84.12 & 89.02       \\ \hline
\multirow{2}{*}{AISHELL-4}  & $\checkmark$    & 91.75 & 91.32 & \rule{0pt}{2.4ex}\textbf{91.53}       \\
                            & $\times$  & 84.78 & 92.51 & 88.48       \\ \hline
\multirow{2}{*}{DIHARD3}    & $\checkmark$    & 94.16 & 90.36 & \rule{0pt}{2.4ex}\textbf{92.22}       \\
                            & $\times$  & 93.86 & 89.96 & 91.88       \\ \bottomrule
\end{tabular}
\vspace{-0.4cm}
\end{table}

\vspace{-0.1cm}

\subsection{Experiments with artificial database}

To further assess the generalization ability of SCDNet and provide additional references for future work, we trained SCDNet based on artificial data and evaluated the model on different datasets. The results are shown in Table~\ref{tab:artificial}. As demonstrated, in the four test sets, the performance degradation of SCDNet trained with artificial data is within 10\% compared to the model trained directly in the corresponding domain, as shown in Table~\ref{tab:ablation}. This showcases that the proposed SCDNet can generalize to unseen domains when trained solely with artificial data.

\vspace{-0.2cm}
\begin{table}[th]
\caption{Results of SCDNet trained with the artificial data.}
\label{tab:artificial}
\vspace{-0.2cm}
\setlength{\tabcolsep}{4.53mm}
\begin{tabular}{lccc}
\toprule
Dataset    & \multicolumn{1}{c}{Cov(\%)} & \multicolumn{1}{c}{Pur(\%)} & \multicolumn{1}{c}{F1(\%)} \\ \midrule
AMI        & 92.94 & 82.32 & 87.31            \\
AliMeeting & 90.37 & 73.76 & 81.23            \\
AISHELL-4  & 85.72 & 87.29 & 86.50            \\
DIHARD3    & 96.35 & 84.27 & 89.90            \\ \bottomrule
\end{tabular}
\vspace{-0.2cm}
\end{table}

\vspace{-0.1cm}
\section{Conclusions}
In this paper, we introduce a self-supervised learning feature-based SCD model called SCDNet. With SCDNet, various SSL models, including Hubert, wav2vec 2.0, and WavLm, are explored. A weighting fusion strategy is employed to assess the effectiveness of representations from different layers in a pre-trained SSL model. This strategy efficiently identifies a better layer compared to the last layer. Results obtained by SCDNet using different representations indicate the suitability of the representation from layer 3 of the WavLm-base model. Additionally, a fine-tuning-based method is employed to evaluate different SSL models for the SCD task, with the results highlighting the strong performance of WavLm, regardless of the scale. Furthermore, both SCDNet and the fine-tuning-based method outperform previous SOTA results, showcasing the efficacy of SCDNet's design and the effectiveness of the proposed contrastive learning approach.

\bibliographystyle{IEEEtran}
\bibliography{mybib}

\end{document}